\documentclass[11pt,a4paper]{article}
\pdfoutput=1

\usepackage{jheppub,slashed}
\usepackage[dvipsnames]{xcolor}

\hypersetup{bookmarks=true,unicode=true,pdftoolbar=true,pdfmenubar=true,colorlinks=true,
    linkcolor=blue,citecolor=magenta,filecolor=magenta,urlcolor=cyan}

\title{Phenomenological analysis of multi-pseudoscalar mediated dark matter models} 

\author[a]{Shankha Banerjee}
\author[b]{\!\!, Genevi\`eve B\'elanger}
\author[c,d]{\!\!, Disha Bhatia}
\author[e]{\!\!, Benjamin Fuks}
\author[f]{\! and Sreerup Raychaudhuri}

\emailAdd{shankha.banerjee@cern.ch}
\emailAdd{belanger@lapth.cnrs.fr}
\emailAdd{dishabhatia@imsc.res.in}
\emailAdd{fuks@lpthe.jussieu.fr}
\emailAdd{sreerup@theory.tifr.res.in}

\affiliation[a]{Theoretical Physics Department, CERN,
Esplanade des Particules 1, Geneva CH-1211, Switzerland}
\affiliation[b]{LAPTh, CNRS, USMB,
9 Chemin de Bellevue, Annecy 74940, France}
\affiliation[c]{Institute of Mathematical Sciences, CIT Campus, Taramani, Chennai 600113, India }
\affiliation[d]{Homi Bhabha National Institute, BARC Training School Complex,
Anushakti Nagar, Mumbai 400094, India}
\affiliation[e]{ Laboratoire de Physique Th\'eorique et Hautes \'Energies,  UMR 7589, Sorbonne Universit\'e et CNRS, 4 place Jussieu, Paris 75252 Cedex 05, France}
\affiliation[f]{Tata Institute of Fundamental Research, Mumbai 400005, India}

\abstract
{Non-minimal simplified extensions of the Standard Model have gained considerable currency in the context of dark matter searches at the LHC, since they predict enhanced mono-Higgs and mono-$W/Z$ signatures over large parts of the parameter space. However, these non-minimal models obviously lack the simplicity and directness of the original simplified models, and are more heavily dependent on the model assumptions. We propose to classify these models generically on the basis of additional mediator(s) and dark matter particles. As an example, we take up a scenario involving multiple pseudoscalar mediators, and a single Dirac dark matter particle, the latter being a popular introduction to ensure ultraviolet completion of theories with multiple pseudoscalar fields. In the chosen scenario, we discuss the viable channels and signatures of relevance at the future runs of the LHC. These are then compared with the minimal simplified scenarios and distinguishing features are pinpointed.
}

\begin{document}

\begin{flushright} CERN-TH-2021-174 \end{flushright}

\maketitle
\flushbottom

\section{Introduction}
Weakly-interacting massive particles (WIMPs) with masses around the electroweak scale are popular candidates \cite{Steigman:1984ac,Fox:2019bgz} to explain the observed dark matter (DM) relic density \cite{Planck:2018vyg}. Following the so-called `WIMP miracle' paradigm, they appear quite naturally  in several popular extensions of the Standard Model (SM) which try to address the hierarchy problem~\cite{Hooper:2009zm}. They have the added advantage that they can be probed experimentally by studying their elastic scattering with SM particles (DM direct detection) \cite{Goodman:1984dc}, by analysing their annihilation into SM particles (DM indirect detection) \cite{Gaskins:2016cha} or by producing them in the collisions of SM particles at high-energy accelerators~\cite{Boveia:2018yeb}. These different methods probe different energy scales and hence provide complementary information on the existence -- or otherwise -- of WIMPs. Unfortunately, the current results from such experiments yield a uniformly null outcome, as a result of which all that we possess are fairly severe constraints on the parameter space of the more well-motivated models in which WIMPs are embedded \cite{Fermi-LAT:2015att,LUX:2017ree,Aprile:2018dbl,Boveia:2018yeb,PandaX-4T:2021bab}. This is all the more reason, therefore, to approach the dark matter problem in a model-independent way, using the framework of either effective field theories (EFTs) \cite{Fitzpatrick:2012ix,Criado:2021trs} or simplified models \cite{Abdallah:2014hon,Abercrombie:2015wmb,Arina:2020udz}. Here, we can then try to pinpoint the specific Lorentz structures which are permitted by the existing constraints, rather than delving deep into the intricacies of a specific ultraviolet (UV)-complete model~\cite{Ellis:1983ew}.

Of the two above-mentioned frameworks to develop a model-independent study of dark matter, EFTs possess the virtue of consistency under evolution of the model at different energy scales. However, they are generally plagued with a proliferation of undetermined constants and new higher-dimension operators of unknown origin, making a phenomenological analysis diffuse, and their UV completion may be questionable when new states are light. On the other hand, simplified models are easier to analyse, having a smaller number of new particles and interactions to consider than a typical UV-complete model. At the same time, they embed a simple structure in terms of dimension-four operators with a clear physical meaning assignable to the unknown parameters of the model. Thus simplified 
models, even if they represent a less complete picture of nature than EFTs or UV-complete models, are much more amenable to a phenomenological analysis. This is the primary motivation to use this kind of framework to study dark matter interactions. 

Typical simplified models for dark matter~\cite{Abdallah:2014hon,Abercrombie:2015wmb,Arina:2020udz} include, in addition to the SM particle content, one dark matter particle $\chi$ which can be of any spin and one mediator particle $X$. In the $s$-channel configuration (on which we focus in this study), the mediator $X$ is generally taken to be a scalar or a vector. Such a model has four new parameters, namely the masses of the new fields and the couplings of the mediator with the dark matter and the SM particles, the latter being usually assumed universal.

Simplified models such as the above construction are essentially toy models which have proved extremely useful in providing generic trends in dark matter searches and in helping to understand how to relate the different kinds of search results available. For example, comparison of constraints from the dark matter relic density and those from direct/indirect searches is straightforward. In fact, when this is done, the combined constraints are quite stringent~\cite{Arcadi:2017kky,Arina:2020tuw}. Cross sections for dark matter interactions as low as those demanded by direct detection data would lead, at the cosmological scale, to an early freeze-out and hence a high relic density which would over-close the Universe. In order to evade such an impasse, we require a model where the cross sections in the early Universe are much greater than at the present epoch. One simple way to do this is to assume a mediator mass such that the dark matter co-annihilation cross sections in the early Universe would be of a resonant nature, and therefore large. When the temperature falls to its present value, we will be far away from the resonance and the cross section will get reduced to a minuscule fraction of the resonant corresponding value. Another possibility, that is an attractive one, is that the mediator field be of pseudoscalar nature. In this case the relevant cross sections can be shown to be velocity-suppressed~\cite{Agrawal:2010fh,Berlin:2014tja}. Obviously, given the drop in kinetic energy from the early Universe, the cross sections in the present epoch will be very small. Thus, even within simplified models, one must make specific model assumptions to advocate phenomenological viability.

Simplified models of the type described above can often lead to unitarity violation at fairly low energy scales, even before that one reaches the scale at which a new, presumably UV-complete, model takes over. This was explicitly shown, for instance, in ref.~\cite{Kahlhoefer:2015bea} for a $Z'$-mediated DM model and in refs.~\cite{Fuks:2020tam,Baldes:2021aph} for dilaton-assisted DM (where dimension-five operators are in order). On general grounds, this is also easy to understand since simplified models, which just add a minimal particle content to the SM, are almost certainly incomplete theories. Unitarity violation is obviously a hallmark of such incomplete theories. Another deficiency is that at a collider like the LHC, a given simplified model may not be able to predict many of the viable signals for dark matter. At a $pp$ collider, in fact, the only really observable process in a simplified model would be $pp \to X^* j \to \chi\bar{\chi} j$, where the $X^*$ state stands for an off-shell mediator and the jet $j$ arises from initial-state radiation. This will lead to an enhancement in the mono-jet production rate ({\it i.e.} in the rate of events featuring an energetic jet and missing transverse energy $\slashed{E}_T$), by means of which the dark matter might be detected at the LHC~\footnote{A corresponding signature with an initial-state photon radiation would also be detectable as a monophoton signal. Its rate is however much suppressed compared to the mono-jet one.}. However, there may be other signals for dark matter. As an example, let us assume that the underlying UV-complete theory (whatever it is) contains two Higgs doublets, as well as a pseudoscalar singlet. After electroweak symmetry-breaking and mixing, then, there will be two physical pseudoscalars $P_1$ and $P_2$, with (say) $P_2$ being much more massive than $P_1$. In such a case, we might have a process such as $pp \to P_2^* \to H + P_1 \to H + \chi\bar{\chi}$, where $H$ stands for the SM 125~GeV Higgs scalar. The corresponding final signal will consist of a Higgs boson and missing energy, which would not be predicted in the usual simplified models at all. Similarly, one could also consider a $Z+\slashed{E}_T$ signal if there is a $ZP_{1/2} H$ coupling.

For quite some time, therefore, it has been pointed out \cite{Bell:2015sza,Kahlhoefer:2015bea,Englert:2016joy,Goncalves:2016iyg,Albert:2016osu,Morgante:2018tiq,Cornell:2021crh} that simplified models may, in fact, be over-simplified. It is thus reasonable to examine so-called `less-simplified models', where the dark sector includes more than one field $\chi_i$ ($i=1,2,\dots$) and/or where there are more than one mediator $X_j$ ($j=1,2,\dots$). Obviously, this leads to a proliferation of new operators and unknown couplings, since such extensions have also to be considered in a model-independent way. In other words, we ignore any underlying symmetry principles, such as, for example, those that would be assumed in a UV-complete theory which would be relevant at high energies. However, it is easy to adjust the parameters to push the scale of unitary violation to much higher energies than in minimal simplified models, so that this scale becomes comparable with what we would get in an EFT approach. It is also clear that some more -- if not all -- of the possible collider signals will be predicted in less-simplified models. For example, if a less-simplified model has two pseudoscalar mediators $X_1$ and $X_2$ with $X_1$ significantly more massive than $X_2$, and if there exists an $HX_1X_2$ coupling, then it will be easy to generate not just a mono-jet signal through $pp \to X^*_{1,2}j \to \chi_i\bar{\chi}_j j$ (with $i,j = 1,2$), but also a mono-Higgs signal through $pp \to X_1^* \to H + X_2 \to H + \chi_i\bar{\chi}_j$, as well as the analogous $Z+\slashed{E}_T$ signal. For these reasons, less-simplified models have been an increasingly popular approach in recent times~\cite{Morgante:2018tiq,Cornell:2021crh}.

\begin{figure}
\begin{center} 
\includegraphics[width=0.75\textwidth]{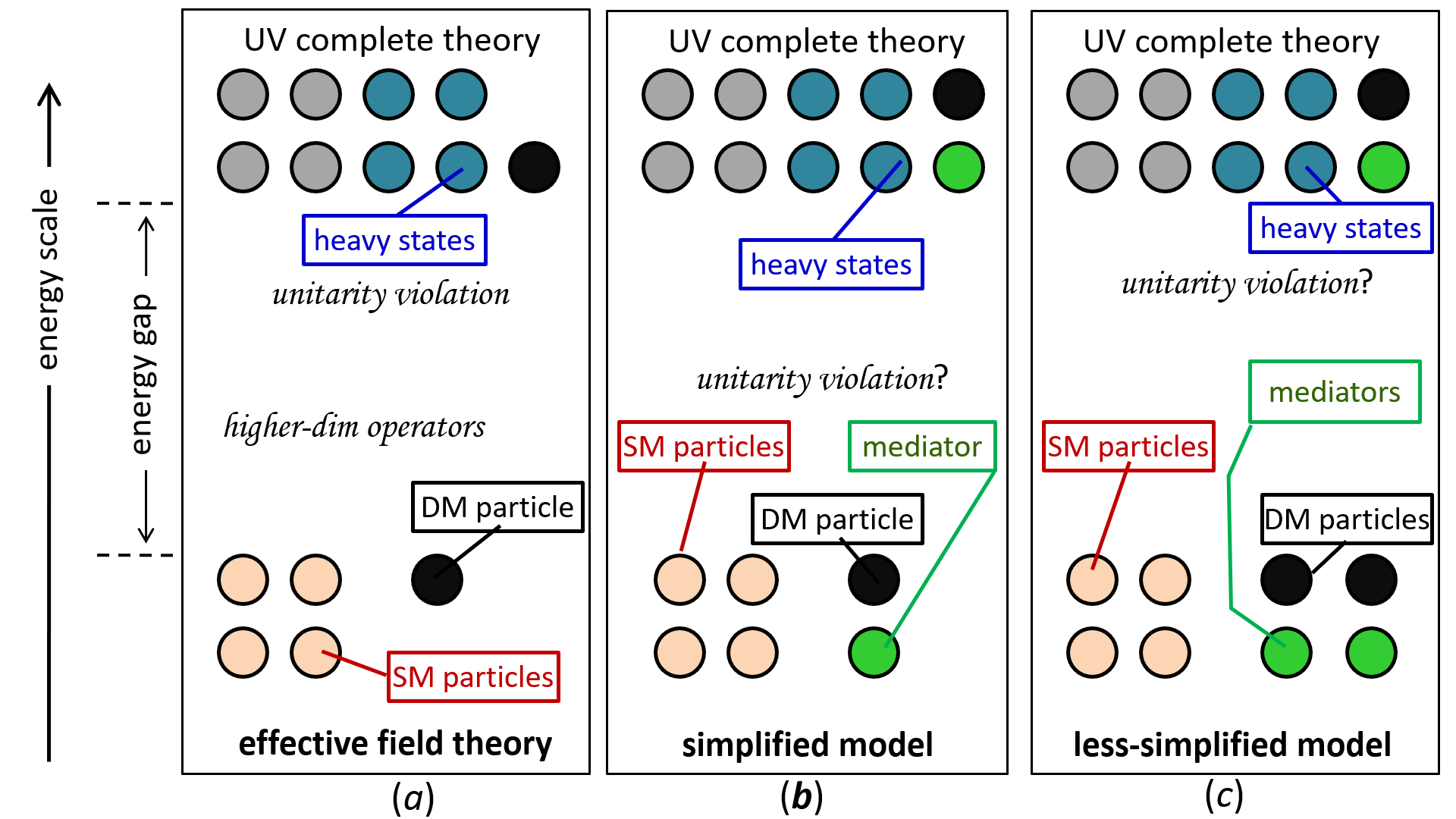}
\caption{ \label{fig1}We compare the three most commonly used frameworks to study dark matter at colliders. These consist of EFTs (left), simplified models (centre) and less-simplified models (right). The pink circles represent the SM particles, the black circles the dark matter particles and the green circles the mediators. At the UV scale, the grey circles are associated with the SM fields without electroweak mixing and the teal circles with the heavy states, whereas the DM particles and the mediators are still represented by black and green circles respectively.}
\label{fig:frameworks}
\end{center}
\end{figure}
The above frameworks for studying dark matter without referring to a complicated multi-parameter UV-complete extension of the SM are summed up in figure~\ref{fig:frameworks}. The panels, from left to right represent, respectively, ($a$) EFTs, ($b$) simplified models and ($c$) less-simplified models. At the lower end of each panel, the pink circles represent the SM particle content, the black circles represent the dark matter content and the green circles represent the mediators. At the upper end of each panel, the underlying UV-complete theory is represented by grey circles for the SM fields when not accounting for electroweak-scale mixing, teal circles for the new heavy states, and the dark matter and mediator fields appear as before through black and green circles respectively. As mentioned above, unitarity may be violated in all three classes of models, but it lies at a comparatively high scale in panels ($a$) and ($c$), whereas it is associated with a rather low scale in (b). Other schematic figures may be found elsewhere in the literature~\cite{Morgante:2018tiq}, with more or less similar features.

At this point, it may be stated that the only serious phenomenological studies of less-simplified frameworks undertaken so far have been in the two-Higgs doublet plus pseudoscalar model~\cite{Bauer:2017ota,LHCDarkMatterWorkingGroup:2018ufk}. While useful in ways described above, taking a specific model lacks generality and, therefore, militates against the spirit of simplified models. In this article, therefore, we attempt a systematic model-independent study of less-simplified models with {\it two} pseudoscalar mediators, and different kind of particle content at low energies. We thus have two purposes. First we aim to perform an analysis in the less-simplified model framework in a model-independent language, where particles are characterised by their masses, spins and couplings without reference to any underlying symmetries. Next we plan to derive updated constraints on the model parameters from current experimental data.

Though the above may sound rather ambitious, the actual philosophy adopted in this work is rather simple-minded. We categorise less-simplified models on the basis of the additional number of particles required to explain their dark matter phenomenology. We then motivate the case for having pseudoscalar mediators, which has already been touched on above. This is followed by an analysis of the effect of having extra low-energy states. Our analysis thus easily generalises the two-Higgs doublet plus pseudoscalar model to a case when the same particle content is embedded in a larger model. We have checked, however, that when we impose the same constraints on the parameters as are present in that specific theory, our results closely tally with those of refs.~\cite{Bauer:2017ota,LHCDarkMatterWorkingGroup:2018ufk}.

This article is organised as follows. Section~\ref{sec:models} discusses minimal simplified models for dark matter and outlines the generic form of a less-simplified model. In particular, we describe the two-pseudoscalar-mediated less-simplified model that we investigate. Section~\ref{sec:exp} details the experimental constraints which we have used in our analysis and section~\ref{sec:constraints} summarises our updated findings of the allowed parameter space. Our conclusions are set forth in section~\ref{sec:conclusions}.
 
%%%%%%%%%%%%%%%%%%%%%%%%%%%%%%%%%%%%%%%%%%%%%%%%%%%%%%%%%%%%%%
\section{Simplified models for pseudoscalar mediated DM}
\label{sec:models}
%%%%%%%%%%%%%%%%%%%%%%%%%%%%%%%%%%%%%%%%%%%%%%%%%%%%%%%%%%%%%%%%%%%%%%%%%%%%%%%%%
\subsection{From single-pseudoscalar to multi-pseudoscalar mediated dark matter}
\label{sec:1simp}
%%%%%%%%%%%%%%%%%%%%%%%%%%%%%%%%%%%%%%%%%%%%%%%%%%%%%%%%%%%%%%%%%%%%%%%%%%%%%%%%
We begin by providing details on the minimal pseudoscalar-mediated fermionic dark matter model. This is one of the simplest and well-motivated simplified models for dark matter, owing to the negligible constraints arising from direct detection. In this setup, a dark matter candidate $\chi$ of mass $m_\chi$ and a pseudoscalar mediator $P$ of mass $m_P$ are introduced on top of the SM degrees of freedom, with the assumption that any other field beyond the SM (BSM) remains decoupled. To obtain the most stringent collider constraints on the model, it is assumed that the pseudoscalar interacts dominantly with the SM quarks and DM. For simplicity, we neglect its couplings with leptons. Interactions with quarks are enforced to flow from minimal flavour violation~\cite{DAmbrosio:2002vsn}. They hence possess a structure implying that the interactions with a given quark flavour $q$ are suppressed by $m_q/v$, where $m_q$ is the quark mass and $v$ is the vacuum expectation value of the Higgs field. Furthermore, interactions among different quark flavours are considered to be vanishingly small, which helps to relax the stringent constraints originating from neutral meson mixing data.

The corresponding interaction Lagrangian $ \mathcal{L}_{\text{int},1}$ is parametrised as
\begin{equation}\begin{split}
  \mathcal{L}_{\text{int},1} =&
    - \sum_q \left(\frac{i y_q g_q}{\sqrt{2}}\  \bar{q} \gamma_5 q\; P \right)
    - i y_{\chi} \bar{\chi} \gamma_5 \chi \; P
    + A_{P}\;  F_{\mu\nu} \widetilde{F}^{\mu\nu} \;  P
    + G_{P}\; G_{a,\mu\nu} \widetilde{G}^{a,\mu\nu} \; P.
\end{split}\label{eqn:simp1}\end{equation}
In this expression, $y_q = m_q/v$, $g_q$ is the coupling constant modifier for a given quark species $q$, and $y_\chi$ is the dark matter coupling with the pseudoscalar. Moreover, the effective couplings of the pseudoscalar with photons and gluons $A_P$ and $G_P$ are functions of the Yukawa couplings themselves~\cite{Djouadi:2005gj}. In our notation the subscript `1' in the Lagrangian symbol $\mathcal{L}_{\text{int},1}$ reflects that we consider a single-mediator scenario, which thus comprises the free parameters,
\begin{equation} 
\bigg\{ g_q,\; y_\chi,\; m_P,\; m_\chi \bigg\}.
\end{equation}

As mentioned above, this model is among the simplest simplified model for dark matter. In the context of a UV-complete picture, we impose that all BSM particles different from the pseudoscalar mediator and that could be present in the spectrum are decoupled. However, this assumption may not always be a reasonable one. Heavier but not decoupled mediators may in some cases play a crucial role to detect dark matter signals at colliders, in particular in channels other than the mono-jet ones. They may even correspondingly yield more stringent constraints on the model than what could be extracted from mono-jet data alone. A consistent understanding of the impact of not too heavy generic mediators which may be present in the full UV-complete theory is therefore important.

A further motivation for such a study can be found by realising that the pseudoscalar interactions with the SM quarks introduced in eq.~\eqref{eqn:simp1} violate electroweak gauge invariance. Promoting such a Lagrangian to an electroweak gauge-invariant one implies that the pseudoscalar must be embedded in some multiplet of $SU(2)_L$, the most common example being a doublet. Moreover, as dark matter is an electroweak singlet, the coupling of dark matter to the mediator necessarily involves an additional pseudoscalar, like in the so-called Two Higgs Doublet + Pseudoscalar Model (2HDM+Pseudoscalar)~\cite{Bauer:2017ota,LHCDarkMatterWorkingGroup:2018ufk}. Consequently, any generic SM extension dealing with singlet dark matter interacting through pseudoscalar mediation should include two pseudoscalars in its particle spectrum. When additional $CP$-even scalars are present, direct detection constraints enforce their effective decoupling. We therefore assume that $CP$-even scalars do not contribute significantly to DM formation in the early Universe, and we consider that the effects of one or more pseudoscalars are quintessential.

In the model-independent analysis which follows, we largely consider that the $CP$-even degrees of freedom remain decoupled, and we focus on an extension of the minimal simplified pseudoscalar-mediated DM model in which a second pseudoscalar is included. We denote this new model, that we present in the following subsection, as a two-pseudoscalar-mediated simplified model.

%%%%%%%%%%%%%%%%%%%%%%%%%%%%%%%%%%%%%%%%%%%%%%%%%%%%%%%%%%%%%%%%%%%%%%%%%%%%%%%%%%%%%%%%%%%%%%%%%%%
\subsection{Two-pseudoscalar-mediated simplified dark matter model}
\label{sec:2simp}
%%%%%%%%%%%%%%%%%%%%%%%%%%%%%%%%%%%%%%%%%%%%%%%%%%%%%%%%%%%%%%%%%%%%%%%%%%%%%%%%%%%%%%%%%%%%%%%%%%%
With the aim of dealing with simplified pseudoscalar-mediated dark matter scenarios in as much of a model-independent manner as we can, we write the corresponding Lagrangian in the mass basis, as done for $Z^\prime$-mediated dark matter in ref.~\cite{Duerr:2016tmh} (but for a model-dependent setup) and in ref.~\cite{Cline:2015qha}, for multi-higgs mediated models. We consider a fermionic DM candidate $\chi$ and two pseudoscalar states $P_1^0$ and $P_2^0$. Their mixing matrix is generically parametrised through a mixing angle $\theta$, so that the mass eigenstates $P_1$ and $P_2$ are defined through
\begin{equation}
  \begin{pmatrix}
    P_1^0 \\
    P_2^0
  \end{pmatrix} = 
  \begin{pmatrix}
  \cos\theta & -\sin{\theta} \\
  \sin\theta & \cos\theta 
  \end{pmatrix}
   \begin{pmatrix}
    P_1 \\
    P_2
  \end{pmatrix}.
\label{eq:Pmix}\end{equation}
As a convention, we assume that $P_2$ is  heavier than $P_1$, {\it i.e.} that the two pseudoscalar masses satisfy $m_{P_2} \geq m_{P_1}$. Moreover, we take $P_1^0$ as interacting with the SM sector only, and $P_2^0$ as interacting with the dark sector only. For the sake of remaining model-independent, we do not specify the details of the underlying UV physics. We instead assume that the interactions of the pseudoscalar $P_1^0$ with the SM quarks, as well as those of the pseudoscalar $P_2^0$ with dark matter, arise in a gauge-invariant manner. We then restrict the Lagrangian to only its relevant terms,
\begin{equation}
  \mathcal{L}^{(0)} \supset
     -\sum_q \left( \frac{i y_q g_q}{\sqrt{2}} \bar{q} \gamma_5 q\; P_1^0\right) 
    - i y_{\chi}\  \bar{\chi} \gamma_5 \chi \; P_2^0 \;.
\end{equation}
As we have written only a subset of all possible Lagrangian terms, the interactions of the SM quarks with the pseudoscalar $P_1^0$ appear to be violating electroweak gauge invariance. Specifying the omitted terms and incorporating a full UV model would however avoid this. We nevertheless assume that the details of the UV physics are not relevant for the description of the dark matter dynamics at collider energy scales and in the early universe, and we therefore safely neglect them in our approach.

It is the mass mixing~\eqref{eq:Pmix} between the two pseudoscalars which generates all the possible interactions between the two sectors,
\begin{equation}\begin{split}
 \mathcal{L}_{\rm mass} \supset&\ 
  -\sum_q \left( \frac{i y_q g_q}{\sqrt{2}} \cos\theta\ \bar{q} \gamma_5 q\; P_1 \right) 
 - i y_{\chi} \sin\theta\ \bar{\chi} \gamma_5 \chi \; P_1 \\ 
  &\ + \sum_q \left( \frac{i y_q g_q}{\sqrt{2}} \sin\theta \ \bar{q} \gamma_5 q\; P_2 \right)-
  i y_{\chi} \cos\theta\ \bar{\chi} \gamma_5 \chi \; P_2 \;.
\end{split}\end{equation}
Those interactions, and in particular their different sign patterns,  could lead to interesting (destructive) interference effects in varied dark matter processes. Moreover, in the limit of degenerate pseudoscalars that maximally mix, the effective interactions between the SM and the dark sector vanish. As in section~\ref{sec:1simp}, all pseudoscalar couplings to the SM fermions are written by enforcing minimal flavour violation, with the hindsight that they originate from an $SU(2)_L \times U(1)_Y$ invariant form of the Lagrangian. Furthermore, we also consider that the pseudoscalars dominantly interact with quarks, and that their interactions do not induce any flavour-changing neutral currents.

Other terms are relevant in the description of the interactions of the two pseudoscalars with the SM. These consist of multi-bosonic interactions given by
\begin{equation}\begin{split}
  \mathcal{L}_{\text{int},2} \supset&\
    A_{P_1}\; F_{\mu\nu}   \widetilde{F}^{\mu\nu}   P_1 +
    A_{P_2}\; F_{\mu\nu}   \widetilde{F}^{\mu\nu}   P_2 +
    G_{P_1}\; G_{a,\mu\nu} \widetilde{G}^{a,\mu\nu} P_1 +
    G_{P_2}\; G_{a,\mu\nu} \widetilde{G}^{a,\mu\nu} P_2 \\
    &\ + m_{11} P_2 P_1 H + m_{22} H P_1 P_1 + m_{33} H P_2 P_2.
\end{split}\label{eqn:simp2}\end{equation}
The subscript `2' in the Lagrangian symbol $\mathcal{L}_{\text{int},2}$ refers to the scenario with two pseudoscalar mediators.
Whereas all the interactions of the pseudoscalars with the SM gauge bosons arise at one-loop and are thus suppressed, we include the interactions with the gluons and the photons that are crucial for collider searches for dark matter. The effective couplings $A_{P_i}$ and $G_{P_i}$ are functions of the Yukawa couplings, and their dominant contributions arise from top and bottom loops. In addition, we include all trilinear couplings of the pseudoscalars with the SM Higgs boson $H$. Implicitly, we have imposed the alignment limit, in which the couplings of the lightest $CP$-even scalar boson $H$ are all SM-like. Such a choice is motivated by current experimental measurements~\cite{CMS:2018uag,higgs-signal-strength-atlas}.

The entire set of free parameters in the two-mediator model is thus given by
\begin{equation}
\bigg\{g_q,\; y_{\chi},\; \theta,\; m_{P_1},\; m_{P_2},\; m_\chi,\; m_{11},\; m_{22},\; m_{33}\bigg\}.
\label{eq:parameters}
\end{equation}

To make our results more accessible, we make some additional simplifying assumptions to minimise the parametric dependence of our analysis. First, the coupling constant modifiers $g_q$ are assumed to be the same across all generations and for up-type and down-type quarks, as traditionally done in the class of DM simplified models studied at colliders~\cite{Abercrombie:2015wmb,Arina:2020udz}. Second, the mixing of the two pseudoscalars is taken maximal, {\it i.e.} $\theta = \pi/4$. Such an assumption will help us to establish the impact of the second pseudoscalar relative to the first one. Third, as our collider analysis does not heavily depend on the choice of $y_\chi$ (as it enters only through the branching ratios of the pseudoscalars), we fix it to a conservative value $y_\chi =1$. This choice is similar to the one considered in ref.~\cite{Boveia:2018yeb}.

We finally reduce the number of free parameters by adopting conservative values for the trilinear scalar interaction strengths. Since we want to deal with this as much model-independently as possible, we ignore any reference to any UV completion and treat these couplings as free parameters. We impose the pseudoscalars to be narrow states, with a width-over-mass ratio satisfying {\color{Black} $\Gamma/M \lesssim 15\%$}\footnote{{\color{Black}There is no consensus on the definition of broad and narrow states, the threshold on the $\Gamma/M$ ratio varying from one author to another. We follow the analysis of Ref.~\cite{Deandrea:2021vje} in which large width effects on a signal and its interference with the SM background start to play a role for width-over-mass ratios of about 10\%--15\%. In the case of a broad state, any signal process should include the decay of this state, and the total new physics contribution then consists of the sum of the signal-only diagrams squared and their non-negligible interference with the Standard Model background (that comprises diagrams featuring the same final state but no new particle beyond the Standard Model). This could lead to different peak and dip structures in some kinematic distributions, and hence potentially affect the sensitivity, in one way or the other, of the LHC analyses considered. We leave a detailed analysis of those effects for future work.}}, and we enforce that the constraints ensuing from Higgs invisible decay searches are fulfilled. In the limit of a $P_2$ width dominated by the $P_2\to P_1 H$ channel, as stemming from the above assumptions when $m_{P_2} \gg m_{P_1}, m_{H}$, we get
\begin{equation}
  \frac{\Gamma(P_2)}{m_{P_2}} = \frac{m_{11}^2 \sqrt{m_{H}^4 + \big(m_{P_1}^2 - m_{P_2}^2\big)^2 - 2 m_{H}^2 \big( m_{P_1}^2 + m_{P_2}^2\big)}}{16 \pi m_{P_2}^4} < 0.1
  \ \ \Leftrightarrow\ \
m_{11} \lesssim 2 m_{P_2}.
\end{equation} 
We choose  $m_{11} = 1$~TeV in our analysis, which is satisfied for the most important regions of the parameter space examined in section~\ref{sec:constraints}. On the other hand, the trilinear couplings $m_{22}$ and $m_{33}$ get stringently constrained from invisible Higgs decays~\cite{higgs-signal-strength-atlas}, which is of course only relevant when $m_{H} > m_{P_1}, m_{P_2}$. Assuming that the pseudoscalars dominantly decay into dark matter, the relevant Higgs branching ratios read
\begin{equation}
  {\rm BR}(H \to P_1 P_1) = \dfrac{m_{22}^2 \sqrt{m_{H}^2 - 4  m_{P_1}^2}}{8\pi m_{H}^2 \Gamma(H)} , \qquad
  {\rm BR}(H \to P_2 P_2) = \dfrac{m_{33}^2 \sqrt{m_{H}^2 - 4  m_{P_2}^2}}{8\pi m_{H}^2 \Gamma(H)}\;.
\end{equation}
To satisfy the experimental bounds, we require that $m_{22}$ and $m_{33}$ are smaller than 2~GeV. These bounds on the $m_{22}$ and $m_{33}$ parameters are additionally kept also when $m_{P_1}, m_{P_2} \geq m_{H}/2$ for simplicity, and as those couplings do not lead to any important modifications of the phenomenology of the model. The involved pseudoscalar fields being identical in the corresponding Lagrangian terms ($m_{22} H P_1 P_1$ and $m_{33} H P_2 P_2$), they are indeed disconnected from any potential resonant enhancement such as those emphasised below.

With all the above assumptions, the set of independent parameters in the two-mediator model is reduced to
\begin{equation}
 \bigg\{ g_q,\;   m_{P_1},\;  m_{P_2},\; m_\chi\bigg\}.
\end{equation}
\textcolor{Black}{We estimate the impact of the adopted simplifying assumptions after the presentation of our main results. To this aim we will investigate the variation of the other parameters appearing in Eq.~\eqref{eq:parameters} one by one.}

%%%%%%%%%%%%%%%%%%%%%%%%%%%%%%%%%%%%%%%%%%%%%%%%%%%%%%%%%%%%%%%%%%%%%%%%%%%%%%%%%%%%%%
\section{Experimental constraints}
\label{sec:exp}
%%%%%%%%%%%%%%%%%%%%%%%%%%%%%%%%%%%%%%%%%%%%%%%%%%%%%%%%%%%%%%%%%%%%%%%%%%%%%%%%%%%%%%%
In this section, we discuss the experimental constraints considered in our analysis of the multi-mediated pseudoscalar dark matter simplified models. 
%%%%%%%%%%%%%%%%%%%%%%%%%%%%%%%%%%%%%%%%%%%%%%%%%%%%%%%%%%%%%%%%%%%%%%%%%%%%%%%%%%%
\subsection{Dark matter aspects}
\label{sec:cosmo}
%%%%%%%%%%%%%%%%%%%%%%%%%%%%%%%%%%%%%%%%%%%%%%%%%%%%%%%%%%%%%%%%%%%%%%%%%%%%%%%%%%%
\noindent
{\bf Relic density:} The dark matter relic density has been extracted  very precisely from measurements of the cosmic microwave background anisotropies
 by the PLANCK collaboration~\cite{Planck:2018vyg},
 
\begin{equation}
\Omega h^2= 0.120 \pm 0.001.
\end{equation}
To determine the regions of the parameter space compatible with such a measurement, we add to this result a theoretical uncertainty related to higher-order corrections of the order of 10\%. Thus we consider the $2\sigma$ range, $ 0.096 < \Omega h^2 <0.144$. It has indeed been shown that higher-order corrections are typically of about 10\%, and can even be larger in certain scenarios. For example, such large corrections are found both in the Minimal Supersymmetric Standard Model~\cite{Baro:2007em,Baro:2009na} and in the Inert Doublet Model~\cite{Banerjee:2016vrp, Banerjee:2019luv, Banerjee:2021oxc, Banerjee:2021anv, Banerjee:2021xdp, Banerjee:2021hal}. In the present work, the relic density is computed at leading order using \textsc{micrOMEGAs 5.1}~\cite{Belanger:2018mqt, Belanger:2013oya}. The dominant channels contributing to DM annihilation consist of annihilations into fermion pairs through the $s$-channel exchange of one of the pseudoscalars, or into pairs of light pseudoscalars when $m_{P_1} < m_{\chi}$. The former process is clearly enhanced near one of the pseudoscalar resonances while it can be suppressed due to a destructive interference between the diagrams featuring $P_1$ and $P_2$ exchanges. This interference depends on both the relative masses of the two pseudoscalars and their couplings to the SM and the DM.\\

\noindent
{\bf Direct Detection:} The direct detection rate when the DM interactions are mediated by a pseudoscalar vanishes in the limit of zero-momentum transfer. Contributions to spin-independent elastic scattering thus occur only at the one-loop level. The associated computations in specific models~\cite{Azevedo:2018exj, Li:2019fnn} have shown that the expected cross section lies below current limits, which are of the order of $\sigma \approx 10^{-11}{\rm pb}$, from XENON1T~\cite{Aprile:2018dbl} and PANDAX-4T~\cite{PandaX-4T:2021bab}. For example, for a fermion dark matter model with a single pseudoscalar mediator~\cite{Li:2019fnn}, {\it i.e.} a model similar to the one considered here, it has been shown that the current experiments are not yet sensitive to ${\cal O}(1)$ couplings. In the following, we therefore ignore constraints ensuing from direct detection. We, however, note that future experiments have the potential to probe the model. In particular, it was shown that when the mediator is light, for example when $m_{P_1}=30\ (10)$~GeV for a DM mass of 50 GeV, a cross section above the neutrino floor is obtained for values of $g_q>0.4\ (0.2)$~\cite{Li:2019fnn}.\\

\noindent	
{\bf Indirect Detection:} Fermi-LAT observations of photons from Dwarf Spheroidal Galaxies (dSph) provide constraints on thermal WIMPs below the electroweak scale. In particular, when DM annihilates into fermion pairs ($b\bar{b}$ or $\tau\tau$), bounds at the 95\% confidence level (CL) rule out thermal cross sections ($\langle \sigma v\rangle = 3\times 10^{-26} {\rm cm}^3/{\rm s}$) associated with WIMPs of mass $m_{\rm DM}\leq 100$~GeV~\cite{Fermi-LAT:2015att}. In our model, annihilations into a $b\bar{b}$ pair dominate. We can thus directly compare the predicted DM annihilation cross section with the 95\% CL exclusions as reported by Fermi-LAT for this channel. These limits can however be somewhat relaxed in analyses that take into account background uncertainties~\cite{Calore:2018sdx}. When $m_{P_1}<m_\chi$, annihilations into a $P_1$ pair are also possible. Such a channel is however $p$-wave suppressed, and is therefore always subdominant.

Searches for DM in anti-proton fluxes as performed by AMS-02~\cite{AMS:2016oqu} can also provide stringent constraints on DM. However, these constraints strongly depend on the cosmic ray (CR) propagation parameters as well as on the DM profile. The uncertainties on the propagation parameters have nevertheless been greatly reduced by AMS-02 measurements of the B/C ratio~\cite{Genolini:2021doh, AMS:2016brs}. We use the limits derived in ref.~\cite{Reinert:2017aga}, when $b\bar{b}$ final states are considered. There, the size of the diffusion halo is fixed to its minimum value $L=4.1$~kpc, and a global fit to B/C and to the antiproton spectrum is performed for the remaining propagation parameters and the exclusion cross section in the $b\bar{b}$ channel for each DM mass. The fit is done for three DM profiles. We use the most conservative limit corresponding to a generalised NFW profile with $\rho=0.3$~GeV/cm$^3$.

\begin{figure}
\begin{center} 
\includegraphics[width=0.45\textwidth]{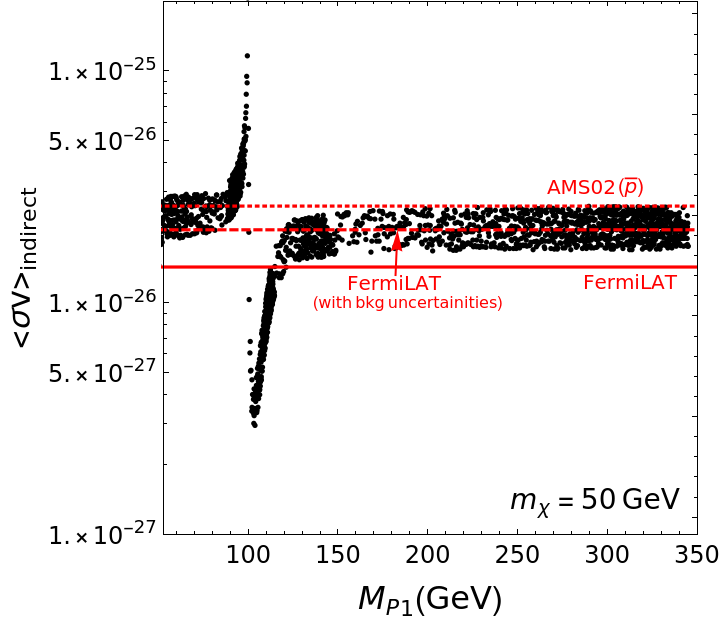}
\includegraphics[width=0.45\textwidth]{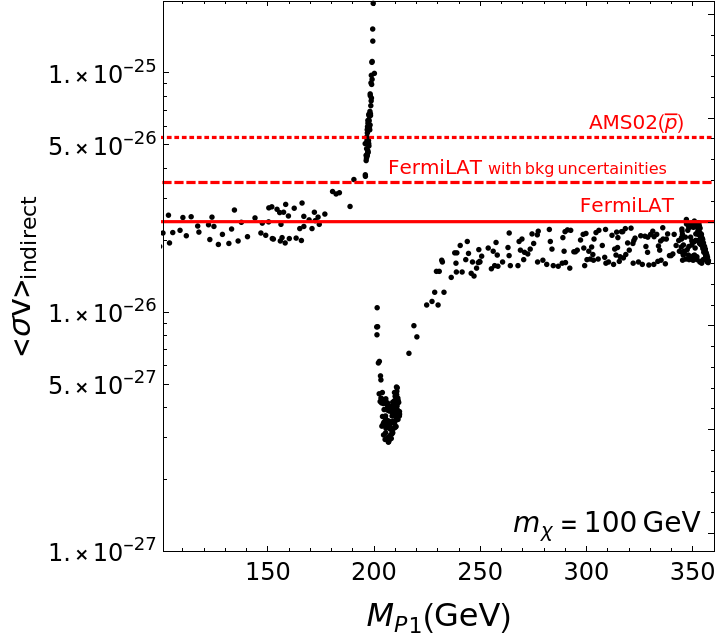}
\caption{\label{fig:2}The behaviour of the indirect detection cross section as a function of the mass of the lightest pseudoscalar $M_{P_1}$. We consider a fixed heavier mediator mass $m_{P_2} = 500$~GeV and dark matter masses of $m_\chi= 50$ GeV (left panel) and $m_\chi=100$ GeV (right panel). In this plot, the couplings of the pseudoscalar with the SM fermions are varied while the couplings to dark matter is fixed to $y_\chi=1$. The constraints from FermiLAT and AMS-02 (anti-protons) are also shown as dashed lines. The regions below the dashed lines are allowed at 95$\%$ CL.} 
\end{center}
\end{figure}

The impact of the dark matter constraints on the model is illustrated in figure~\ref{fig:2}, in which we fix $m_{P2}=500$~GeV, and where we set $m_\chi=50$ GeV (left panel) and $m_\chi=100$ GeV (right panel). Moreover, the coupling $g_q$ is left free and scanned over. In this figure, we display the annihilation cross section $\langle\sigma v\rangle_{\rm ID}=\langle\sigma v\rangle|_{v=0.001c}$ for  all points that fall within the allowed range for the relic density. We first observe that the dominant process being $s$-wave entails that $\langle\sigma v\rangle_{\rm ID}$ is almost constant away from the region $m_\chi=m_{P_1}/2$, with a value that corresponds to its thermal value. The second observation is that there can be a strong enhancement of $\langle\sigma v\rangle_{\rm ID}$. This effect occurs when DM annihilates near a narrow resonance. This is known as a Breit-Wigner enhancement~\cite{Ibe:2008ye, AlbornozVasquez:2011js}. In this region, the cross section is sensitive to the thermal energy. At small velocities relevant for DM annihilation in galaxies, the annihilation process can be enhanced by the pseudoscalar resonance, while at higher velocities such as representative from the early Universe, the annihilation process takes place mostly above the resonance. Conversely, when $m_{P_1}$ is slightly above $2m_{\chi}$, the resonance enhancement is maximal in the early Universe, while no such enhancement occurs at very small velocities in the galaxy (hence the deep dip in figure~\ref{fig:2}).

When $m_{\chi}=50$~GeV, the strict constraints from Fermi-LAT~\cite{Fermi-LAT:2015att} rule out most of the regions allowed by the relic density, except for a narrow region where $m_{P_1}$ is in the range 100--120~GeV. As mentioned above this corresponds to the region where DM annihilation is enhanced in the early Universe. The more conservative  interpretation of the Fermi-LAT limit, which takes into account theoretical uncertainties, allows for most values of $m_{P_1}$. It however restricts the range of values allowed for $g_q$ to those corresponding to the upper range of the allowed interval by the relic density. The conservative limit from AMS-02 antiprotons only constrains the region where $m_{P_1}<2 m_{\chi}$. For heavier DM, on the other hand, the indirect detection constraints become much weaker. The whole range of $m_{P_1}>2m_{\chi}$ becomes allowed along with a significant portion of the lower range.

%%%%%%%%%%%%%%%%%%%%%%%%%%%%%%%%%%%%%%%%%%%%%%%%%%%%%%%%%%%%%%%%%%%%%%%%%%%
\subsection{Collider constraints}
%%%%%%%%%%%%%%%%%%%%%%%%%%%%%%%%%%%%%%%%%%%%%%%%%%%%%%%%%%%%%%%%%%%%%%%%%%%

The multi-mediated pseudoscalar model considered can be tested at the LHC in a twofold manner, either by probing pseudoscalar signals in which the pseudoscalar decays invisibly, or by focusing on its visible decays into a pair of SM states. Signatures featuring invisible decays contribute to various missing transverse energy signals that are traditionally searched for through the so-called mono-$X$ searches for dark matter at colliders~\cite{Birkedal:2004xn, Feng:2005gj, Bai:2010hh, Fox:2011fx, Andrea:2011ws, Bell:2012rg, Petrov:2013nia, Bai:2012xg}. On the other hand, signatures of the visible pseudoscalar decays cannot be used to probe the dark matter properties of the model directly, but could be used instead to corner the pseudoscalar properties and its couplings to the SM. In the following, we only detail the first class of constraints, as the latter class has been found to have a negligible impact.

In order to evaluate the constraints that originate from mono-$X$ analyses at the LHC, we reinterpret the results of ATLAS and CMS searches for mono-jet and mono-Higgs signals. Moreover, our model encompasses minimal flavour violation, and hence the pseudoscalar couplings to third-generation quarks are enhanced. Pseudoscalar production (followed by a decay into a pair of dark matter states) in association with a top-anti-top pair may then play an important role. We correspondingly recast an LHC search for dark matter production in association with a pair of top quarks to quantitatively assess the relevance of this channel.

We derive the LHC sensitivity to our model by means of the aforementioned missing energy searches.
In practice, we rely on the simulation of three relevant collider signals,
\begin{equation}
 p p \to \chi \bar\chi j \quad \text{with } p_T(j)> 100~{\rm GeV}\, , \qquad
 p p \to \chi \bar\chi H \, , \qquad
 p p \to \chi \bar\chi t \bar t \, ,
\end{equation}
where $H$ is the SM Higgs. We employ a state-of-the-art simulation tool chain, relying on the Monte Carlo event generator {\sc MadGraph5\_aMC@NLO} (MG5\_aMC)~\cite{Alwall:2014hca} for the simulation of the above hard processes. The first two signals are evaluated at the next-to-leading order (NLO) in QCD and the last one, featuring a larger final-state multiplicity, at the leading order (LO). This is achieved by importing into MG5\_aMC a UFO library~\cite{Degrande:2011ua} associated with the model introduced in section~\ref{sec:models}. The latter is generated by means of {\sc FeynRules}~\cite{Christensen:2009jx, Alloul:2013bka}, that we jointly use with NLOCT~\cite{Degrande:2014vpa} and {\sc MoGRe}~\cite{Frixione:2019fxg} for the calculation of the UV counterterms and the finite remainders that are necessary to compute loop integrals in four dimensions (as done in MG5\_aMC). To deal with the decay of heavy final-state particles, we make use of the {\sc MadSpin}~\cite{Artoisenet:2012st} and {\sc MadWidth}~\cite{Alwall:2014bza} packages.

All signal events are then passed through {\sc Pythia 8}~\cite{Sjostrand:2014zea} for parton showering and hadronisation, and the resulting hadron-level events are processed by the fast detector simulator {\sc Delphes}~\cite{deFavereau:2013fsa}, as driven by {\sc MadAnalysis}~5~\cite{Conte:2018vmg}. In this setup, the clustering of hadrons into jets is handled according to the anti-$k_T$ algorithm~\cite{Cacciari:2008gp} as implemented in {\sc FastJet}~\cite{Cacciari:2011ma}. To emulate a realistic detector resolution and particle identification performance, we tune the detector response and the reconstruction performance as detailed in the LHC analyses considered. We recast three analyses of partial LHC run~2 data~\cite{ATLAS:2017bfj, ATLAS:2017uis, CMS:2017jrd}, that are all available from the {\sc MadAnalysis~5} Public Analysis Database~\cite{Dumont:2014tja}. The source codes and associated documentation are available from the {\sc MadAnalysis}~5 dataverse~\cite{DVN/DFQPGU_2021, DVN/KAAYFM_2021, DVN/BQM0T3_2021} and ref.~\cite{Fuks:2018yku}.

The considered experimental searches focus on the analysis of partial run~2 data~\cite{ATLAS:2017bfj,ATLAS:2017uis,CMS:2017jrd}. We have therefore verified that the obtained bounds only mildly change after a naive rescaling to the full run~2 luminosity~\cite{ATLAS:2021kxv,CMS:2021far,ATLAS:2021shl}, such a procedure being known to provide results comparable with those that could be obtained from the recast of updated experimental analyses~\cite{Araz:2019otb}. Instead of indicating bounds corresponding to the full LHC run~2, we show below bounds derived from the nominal luminosity of the searches under consideration. We additionally derive estimates for the high-luminosity phase of the LHC by a naive rescaling of the signal and background expectation for a luminosity of 3~ab$^{-1}$, following the prescription of ref.~\cite{Araz:2019otb}.

We estimate the sensitivity of the LHC through the mono-jet channel by reinterpreting the results of a corresponding ATLAS search in 36.1 fb$^{-1}$ of LHC data~\cite{ATLAS:2017bfj}. This search targets events featuring an energetic jet and a large amount of missing transverse energy. It includes 10 inclusive signal regions in which the amount of missing energy has to be larger than some signal-region-dependent threshold, as well as 10 exclusive signal regions dedicated each to a well-defined missing energy bin. The exclusive region `EM2', that enforces 300~GeV $ < \slashed{E}_T < $ 350~GeV, is the most relevant one for the scenarios studied in this work by virtue of the new physics mass range that we study.

For what concerns the mono-Higgs channel, we recast an ATLAS search for mono-Higgs events in 36.1~fb$^{-1}$ of the LHC data~\cite{ATLAS:2017uis}. We consider the resolved signal region of this analysis, in which the Higgs boson decays into a pair of separated narrow $b$-jets. Moreover, selected events contain some missing transverse energy that satisfies 150~GeV $ < \slashed{E}_T < $ 500~GeV, and that is well separated from the bulk of the hadronic activity.

Finally, the LHC sensitivity to the model from $t\bar t +\slashed{E}_T$ production is assessed by recasting the results of the CMS search for such a new physics signature in 35.9~fb$^{-1}$ of data~\cite{CMS:2017jrd}. Whereas this search targets a supersymmetry-like topology in which the signal originates from the production of a pair of top partners ({\it i.e.} top squarks in supersymmetry) which then decay into a dark matter state ({\it i.e.} a neutralino in the supersymmetric case) and a top quark, it can be reinterpreted in other scenarios, as shown in the CMS publication itself with a top-philic $s$-channel dark matter model~\cite{Arina:2016cqj}. The signal region `SR1' of this analysis, that is defined by an inclusive missing transverse energy selection ($\slashed{E}_T > 200$~GeV) and an exclusive stransverse mass bin (100~GeV $ < M_{T2} < $ 140~GeV), is particularly well suited for the scenarios that we examine.

%%%%%%%%%%%%%%%%%%%%%%%%%%%%%%%%%%%%%%%%%%%%%%%%%%%%%%%%%%%%%%%%%%%%%%%%%%%%%%%%%
\section{Constraining multi-pseudoscalar mediated dark matter}
\label{sec:constraints}
%%%%%%%%%%%%%%%%%%%%%%%%%%%%%%%%%%%%%%%%%%%%%%%%%%%%%%%%%%%%%%%%%%%%%%%%%%%%%%%%%

\begin{table}
\begin{center}
\renewcommand{\arraystretch}{1.3}\setlength\tabcolsep{12pt}
\begin{tabular}{c|c c}
 Scenarios              & Relic density                & LHC phenomenology   \\ \hline
 $m_{P_1} \gg m_{P_2}$  & single-mediator case         & single-mediator case \\
 $m_{P_1} > m_{P_2}$    & single-mediator case         & two-mediator case \\[-.25cm]
                        &                              & (enhanced mono-Higgs rates) \\
 $m_{P_1} \sim m_{P_2}$ &  single-mediator case        & single-mediator case \\[-.25cm]
                     &  (as an effective coupling) & (as an effective coupling)
\end{tabular}
\end{center}
\caption{\label{tab:scenarios} Model description and main phenomenological consequences as a function of the mass splitting between the two pseudoscalars.}
\end{table}

In this section, we derive the current constraints on scenarios with either one or two mediators. This exercise allows for a comparison of the obtained bounds in the two cases, and a good understanding of how the second mediator could have an impact. It turns out that this crucially depends on the mass splitting between the two pseudoscalars, as will be more quantitatively shown in the next subsections. For example, when the second mediator is much heavier than the first one, the model phenomenology can be effectively described in terms of a single mediator (as studied {\it e.g.} in refs.~\cite{Banerjee:2017wxi, Abdallah:2014hon, Cohen:2019wxr,DiazSaez:2021pmg}, for the analysis of a pseudoscalar mediator). Moreover, a single-mediator description could also be used when the two pseudoscalars are approximately mass-degenerate. The effect of the second mediator then appears as an overall effective coupling change. The interesting case corresponds to the intermediate setup in which the second mediator is moderately heavy relative to the first mediator and with large enough couplings to the SM. In this case, its presence would not impact early universe cosmology, but could enhance mono-$X$ expectations at colliders. In particular, one could get an increased mono-Higgs production rate for an appropriately chosen mass splitting, without affecting any other mono-$X$ channel. The above discussion is summarised in Table~\ref{tab:scenarios}.

In order to carry out a comprehensive and quantitative analysis, we divide our study into two parts. First, we focus on understanding the phenomenological consequences of the presence of the second mediator. Next, we estimate how the constraints evolve with the dark matter mass. We examine in particular the corresponding sensitivity of the LHC, since as already discussed in section~\ref{sec:cosmo} the dark matter constraints are generally milder. The LHC thus has the potential to impose complementary bounds. This sharply contrasts with models featuring a Higgs portal or a Higgs-like mediation between the SM and the dark sector, or with scenarios in which a combination of direct detection and relic density constraints severely restricts the allowed regions of the parameter space~\cite{Escudero:2016gzx,Ghorbani:2017jls}.

%%%%%%%%%%%%%%%%%%%%%%%%%%%%%%%%%%%%%%%%%%%%%%%%%%%%%%%%%%%%%%%%%%%%%%%%%%%%%%%%%%%%%%%%%%
\subsection{The impact of the second mediator}
\label{sec:second-mediator}
%%%%%%%%%%%%%%%%%%%%%%%%%%%%%%%%%%%%%%%%%%%%%%%%%%%%%%%%%%%%%%%%%%%%%%%%%%%%%%%%%%%%%%%%%

From the naive discussion above, the limit of a very heavy second mediator should lead to results agreeing with the expectation of a single-mediator model~\cite{Banerjee:2017wxi, Abdallah:2014hon, Cohen:2019wxr}. In order to establish such a statement, we consider certain simplifying assumptions allowing for the mapping of the two configurations. We first recall that the Yukawa coupling modifiers $g_q$ are all equal. Furthermore, we fix the dark matter mass and the dark Yukawa coupling $y_\chi$ to specific values following the conventions of Ref.~\cite{Boveia:2018yeb}. This corresponds to the maximum mixing scenario between the two pseudoscalars and helps us to define the decoupling limit solely in terms of the mass of the second pseudoscalar and independently of the size of the couplings. The idea that underlies a single-mediator model is that any additional particle is decoupled and has no phenomenological consequence at energy scales relevant for colliders. Single-mediator models can then be constrained largely by using mono-jet searches. Therefore, we begin our analysis under the assumption that the BSM contributions to mono-jet production are dominated by the effect of the first mediator (when it can decay into dark matter). We then seek out, in a second step, to determine the limit in which other mono-$X$ signatures, such as the dominant mono-Higgs one~\cite{Bauer:2017ota, LHCDarkMatterWorkingGroup:2018ufk, Argyropoulos:2021sav}, can play a role.

We start by defining the model configurations at which the second mediator can be considered as decoupled in the context of mono-jet production. We demand that the associated contributions to the related cross section are smaller than 10\% of the first mediator contributions. This criterion leads to a constraint on the mass of the second mediator,

\begin{equation}\label{eq:decoupledP2}
\frac{g_q^2 y_\chi^2}{m_{P_2}^2} \leq 0.1\ \frac{g_q^2 y_\chi^2}{m_{P_1}^2} \quad
\text{or} \quad m_{P_2} \geq 3.16\ m_{P_1}  \;,
\end{equation}
provided that both pseudoscalar masses are larger than twice the dark matter mass (so that $P_1$ and $P_2$ both could yield the production of a significant amount of missing energy at the LHC).

\begin{figure}
\begin{center}
\includegraphics[width=0.45\textwidth]{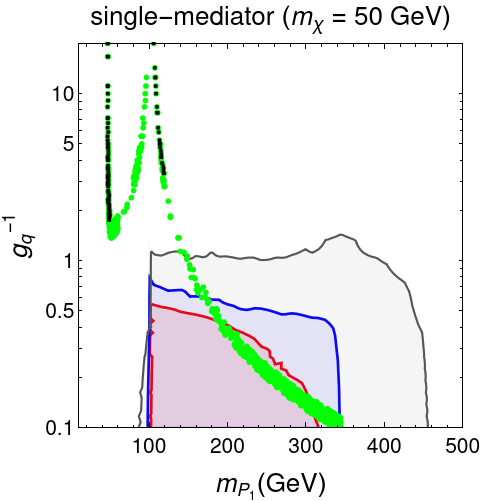}
\includegraphics[width=0.45\textwidth]{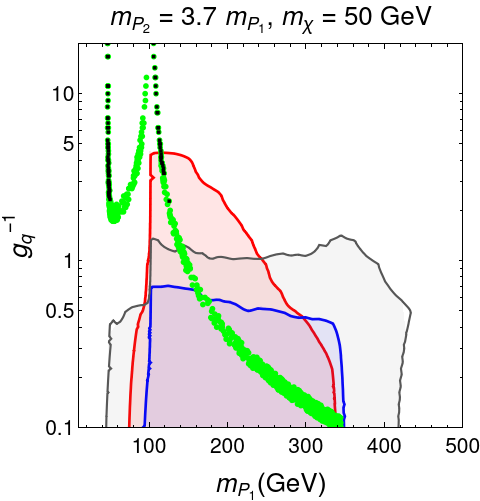}
\caption{\label{fig:comp} Constraints on the model that originate from dark matter relic density, indirect detection and searches at the LHC. The results are shown in the plane $\left( m_{P_1}, g_q^{-1}\right)$ for the single-mediator (left) and the two-mediator (right) case. The points appearing in light green are associated with a relic density satisfying $\Omega h^2 = 0.12 \pm 10 \%$, whereas the dark green points are additionally allowed by the indirect detection results of Fermi-LAT~\cite{Reinert:2017aga}. The shaded regions represent 95\% confidence level exclusions from mono-jet (grey), mono-Higgs (red) and $t\bar{t}+\slashed{E}_T$ (blue) searches in 36~fb$^{-1}$ of LHC data~\cite{ATLAS:2017bfj, ATLAS:2017uis, CMS:2017jrd}.}
\end{center}
\end{figure}

In figure~\ref{fig:comp}, we show constraints on the model parameter space that originate from cosmology and astrophysics as well as from  LHC mono-$X$ searches, both for the single mediator (left) and the two-mediator (right) cases. We find that for a certain range of parameters, mono-Higgs constraints play an important role, even if bounds stemming from mono-jet production are almost identical in the single-mediator and two-mediator scenarios. We recall that the second mediator is enforced to be {\it decoupled in a mono-jet context} as defined in Eq.~\eqref{eq:decoupledP2}.

For a similar reason, $t\bar{t}+\slashed{E}_T$ constraints are not expected to be drastically different between the one-mediator and two-mediator configurations. We indeed find that mono-jet and $t\bar{t}+\slashed{E}_T$ bounds both play a role as soon as the lightest pseudoscalar can decay into dark matter ({\it i.e.} for $m_{P_1}\gtrsim 2 m_\chi$). When the $t\bar t$ threshold is crossed ({\it i.e.} for $m_{P_1}\gtrsim 2 m_t$), $P_1$ decays into a top-anti-top pair dominate and both mono-jet and $t\bar{t}+\slashed{E}_T$ searches lose sensitivity.

In contrast, mono-Higgs production is enhanced in the two-mediator case as soon as the on-shell production of the heavier pseudoscalar $P_2$, followed by a decay into a SM Higgs boson $H$ and a lighter pseudoscalar $P_1$ (that further decays into dark matter), is open,
\begin{equation}
  p p \to P_2 \to P_1 H \to \bar \chi\chi H\, .
\end{equation}
Such a sub-process is important in the mass range considered, its specific impact being related to the analysis cut on the missing energy. The corresponding missing transverse energy spectrum indeed exhibits an edge at~\cite{Bauer:2017ota}
\begin{equation}
  \slashed{E}_{T, {\rm max}} \approx \frac{\lambda\left(m_{P_2}, m_{P_1}, m_H\right)}{2 m_{P_2}} \;,
\end{equation}
where $\lambda(x,y,z)$ stands for the usual K\"all\'en function. As the mono-Higgs analysis that we recast enforces a 150~GeV $ < \slashed{E}_T < $ 500~GeV selection, we can expect the sensitivity of the search to decline for $m_{P_1} \approx 200$~GeV when $m_{P_2} = 3.7 m_{P_1}$.

We now turn to the analysis of the cosmological and astrophysical constraints on the model. The region above the green points in figure~\ref{fig:comp} corresponds to configurations associated with a relic density larger than the observed one. It is thus disallowed. In contrast, for the parameter space region below the green points, the dark matter is under-abundant. Therefore, there should be other sources of dark matter for the model to be phenomenologically viable. We find, as expected, that the value of $g_q$ required is much smaller when $m_\chi\sim m_{P_1}/2$. When $m_\chi>m_{P_1}$ the relic density is instead set by $\chi\chi\to P_1P_1$ annihilations and is therefore independent of $g_q$. Finally, indirect detection bounds are found to be particularly strong for the chosen dark matter mass, and rule out quite a large region of the parameter space.

Dark matter observables hence leave two allowed and disconnected classes of scenarios. In the first one, dark matter cannot be produced via the on-shell decay of the light pseudoscalar ({\it i.e.} $m_{P_1} < 2 m_\chi$). In the second one, we lie close to the resonance ($m_{P_2} \gtrsim 2 m_\chi$) so that Breit-Wigner effects help in reducing $\langle \sigma v\rangle_{\rm ID}$. This consequently allows for avoiding any indirect detection bound close to the resonance peak (see also figure~\ref{fig:2}).

We shall see in the next subsection how these stringent indirect detection constraints can be relaxed by either choosing a heavier DM mass, or by introducing a non-standard cosmological evolution of the early universe.

%%%%%%%%%%%%%%%%%%%%%%%%%%%%%%%%%%%%%%%%%%%%%%%%%%%%%%%%%%%%%%%%%%
\subsection{The impact of the dark matter mass}
\label{sec:dark-matter-mass}
%%%%%%%%%%%%%%%%%%%%%%%%%%%%%%%%%%%%%%%%%%%%%%%%%%%%%%%%%%%%%%%%%%

\begin{figure}
\begin{center} 
\includegraphics[width=0.45\textwidth]{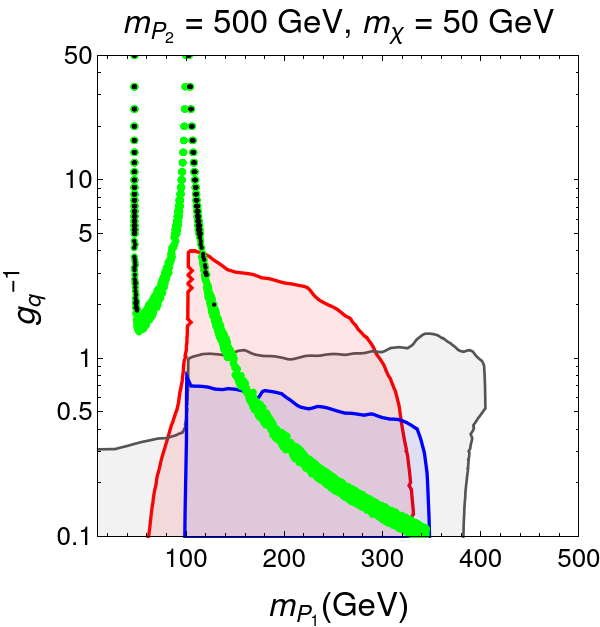}
\includegraphics[width=0.45\textwidth]{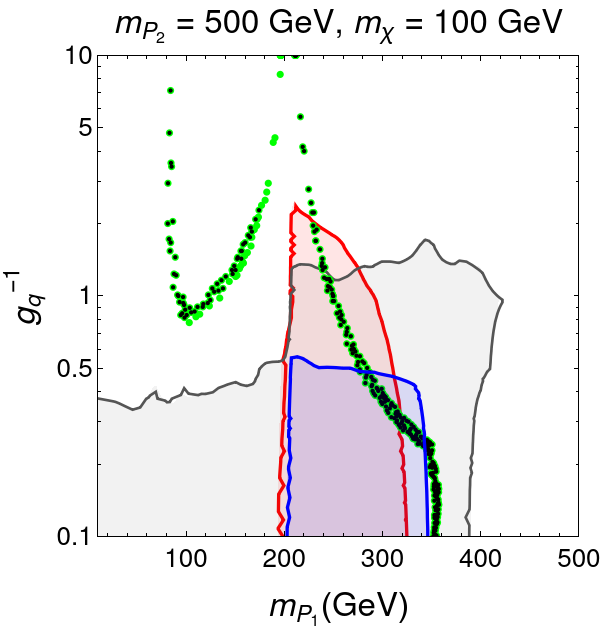}
\caption{\label{fig:mass}Same as in figure~\ref{fig:comp}, but for different mass setups in the two-mediator case. We consider $m_{P_2} = 500$~GeV, together with $m_\chi = 50$~GeV (left) and 100~GeV (right).}
\end{center}
\end{figure}

In this section, we study the evolution of the constraints on the model as a function of the dark matter mass in scenarios where the two mediators play a role. For this purpose, we fix the mass of the second pseudoscalar to $m_{P_2} = 500$~GeV, a value well within the regime in which the second mediator has an impact on the model phenomenology. We then examine two representative classes of benchmark points in which the dark matter mass is fixed to 50 and 100 GeV respectively. The resulting constraints are represented in the $(m_{P_1}, g_q^{-1})$ plane in figure~\ref{fig:mass}.

We first observe that the dark matter mass strongly impacts the cosmology in the model. The relic density curve (in green) on which $\Omega h^2 = 0.12$ is essentially translated towards higher values of $m_{P_1}$, as the resonant regime ({\it i.e.} $m_{P_1} \sim 2 m_\chi$) is correspondingly shifted from $m_{P_1}\sim 100$~GeV to 200~GeV. Indirect detection bounds additionally get  weaker with the increase of the dark matter mass $m_\chi$ (see figure~\ref{fig:2}).

All collider constraints considered also get weaker with increasing values of $m_\chi$. The mono-jet and the $t\bar{t}+\slashed{E_T}$ constraints are irrelevant in regions where the decay of the lightest pseudoscalar into dark matter is closed ($m_{P_1} < 100$~GeV and 200~GeV in the two classes of scenarios under consideration respectively). Moreover  sensitivity is lost when $m_{P_1} \lesssim m_{P_2}$, since in the equal coupling limit, the contributions due to the two pseudoscalars destructively interfere (and cancel out in the equal mass limit). In the intermediate mass configuration ($2m_\chi < m_{P_1} \lesssim m_{P_2}$), $t\bar{t}+\slashed{E_T}$ constraints are still strong and mostly not affected by the change in the dark matter mass, whereas the mono-jet ones get reduced by virtue of the missing energy dependence (or equivalently because of the dark matter mass dependence) of the selection efficiencies in the various signal regions of the analysis. A similar conclusion can be drawn for what concern mono-Higgs constraints that globally weaken with higher mass values of the dark matter. This is due to the shift in the pseudoscalar dark decay threshold (that increases from 100~GeV to 200~GeV when $m_\chi$ varies from 50~GeV to 100~GeV). On the other hand, the bounds in the heavier pseudoscalar $P_1$ regime stay independent of the dark matter mass. In this case, mono-Higgs searches indeed lose sensitivity at approximately $m_{P_1} \approx 320$ GeV for $m_{P_2} = 500$ GeV, regardless of the dark matter mass.

To summarise, we have found that while cosmological constraints on multi-pseudoscalar mediated dark matter models get weaker with an increase in the dark matter mass, the LHC searches still have the potential to probe large regions of the parameter space. \textcolor{Black}{To ascertain the generality of the obtained results, we examine in the next subsections the impact of varying the remaining free parameters of the model. Finally we will close the section by estimating the reach of the high-luminosity phase of the LHC (the HL-LHC) to the model that we investigated in this work.}

%%%%%%%%%%%%%%%%%%%%%%%%%%%%%%%%%%%%%%%%%%%%%%%%%%%%%%%%%%%%%%%%%%%%%%%%%
\subsection{\textcolor{Black}{The impact of the dark matter Yukawa coupling}}
\label{sec:yx}
%%%%%%%%%%%%%%%%%%%%%%%%%%%%%%%%%%%%%%%%%%%%%%%%%%%%%%%%%%%%%%%%%%%%%%%%%
\begin{figure}
\begin{center}
\includegraphics[width=0.43\textwidth]{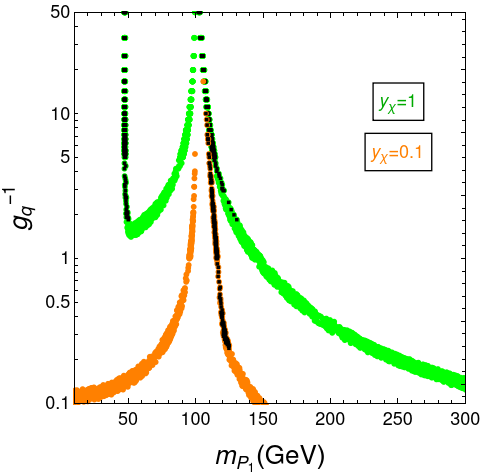}
\includegraphics[width=0.43\textwidth]{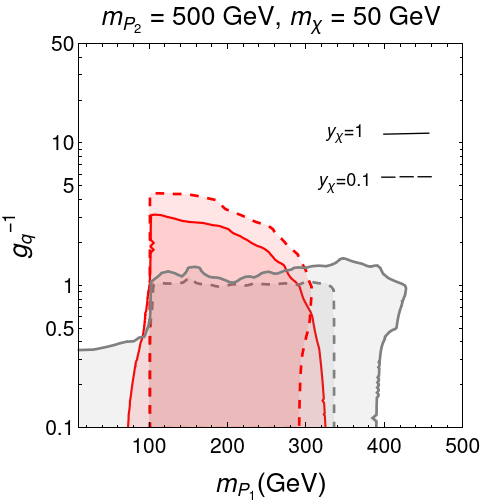}
\caption{\label{fig:yx}\textcolor{Black}{Constraints on the model originating from the dark matter relic density and indirect detection bounds (left), as well as LHC constraints (right), for two representative values of $y_\chi$. The results are displayed in the plane $\left( m_{P_1}, g_q^{-1}\right)$. The points that appear in light green (orange) are associated with a relic density satisfying $\Omega h^2 = 0.12 \pm 10 \%$ for $y_\chi = 1 (0.1)$, whereas the darker points are additionally allowed by indirect detection bounds. The shaded regions in the right panel represent 95\% confidence level  exclusions from mono-jet (grey) and mono-Higgs (red) searches ~\cite{ATLAS:2017bfj, ATLAS:2017uis, CMS:2017jrd}. }The solid (dashed) lines corresponds to exclusions for $y_\chi = 1$ $(0.1)$.}
  \end{center}
  \end{figure}

\textcolor{Black}{In previous sections, we had fixed the value of the DM Yukawa coupling with the pseudoscalars to be one, based on the argument that this choice should have a negligible impact on the extracted collider constraints. This argument is strictly valid only in the case where the pseudoscalar dominantly decays to pair of dark matter particles. On the other hand, the value of  $y_\chi$ should affect cosmological constraints. In this subsection, we therefore analyse the impact of $y_\chi$ on both the cosmological and LHC bounds on the model. The exercise has been performed by fixing the dark matter mass to 50 GeV, the second pseudoscalar mediator mass to 500 GeV, and by keeping  $m_{11}= 1$ TeV,  $m_{22}= m_{33}=2$ GeV, and $\theta=\pi/4$. }

\textcolor{Black}{In figure~\ref{fig:yx}, we  compare the obtained cosmological and LHC constraints for two values of $y_\chi=1$ and 0.1. There are broadly three effects which arise due to the variation of $y_\chi$.
\begin{itemize}
  \item  When $m_\chi< m_{P_1}$, DM dominantly annihilates into a pair of SM particles, $\chi \chi \to$ SM SM, and the annihilation cross section scales as the product of the two Yukawa couplings $g_q^2 y_\chi^2$. In this case, imposing the relic density requirements fixes the cross section to a particular value for a given DM mass, so that a decrease in  $y_\chi$ leads to an increase in  $g_q$. This can be seen on the left panel of figure~\ref{fig:yx}. The minimum allowed value of $g_q$ satisfying the relic density constraint is obtained when $y_\chi$ attains its maximum value set by perturbative unitarity considerations. However, when DM annihilation proceeds near the pseudoscalar resonance, a modification of $y_\chi$ has little impact.
  \item Another interesting feature can be seen when $m_{\chi} \geq M_{P_1}$ with sufficiently large $y_\chi$ values like for instance $y_\chi=1$. The dominant annihilation mode becomes $\chi \chi \to P_1 P_1$ and $g_q$ turns to be unconstrained by cosmology. For smaller values of $y_\chi$  the annihilation cross section to $P_1 P_1$ decreases, and the relative contribution of the SM final states becomes more relevant. This automatically leads to a constraint on $g_q$. This can be clearly seen in the left panel of figure~\ref{fig:yx} for $y_\chi =0.1$.
  \item As expected, we find that the LHC constraints remain largely unaffected when the value of $y_\chi$ decreases, as illustrated in figure~\ref{fig:yx}.  This fully justifies our choice of fixing $y_\chi$ to one in the previous sections. In the results that we presented in the current section, we did not include constraints originating from $t\bar{t}+\slashed{E}_T$ searches, as they were found to be similar and milder than the mono-jet bounds.
\end{itemize}
}

%%%%%%%%%%%%%%%%%%%%%%%%%%%%%%%%%%%%%%%%%%%%%%%%%%%%%%%%%%%%%%%%%%%%%%%%%%%%%%%%%%%%%%%%%%%%%%%
\subsection{\textcolor{Black}{{The impact of the mixing angle $\theta$}}}
%%%%%%%%%%%%%%%%%%%%%%%%%%%%%%%%%%%%%%%%%%%%%%%%%%%%%%%%%%%%%%%%%%%%%%%%%%%%%%%%%%%%%%%%%%%%%%%
\textcolor{Black}{The variable $\theta$ characterises the mass mixing between the two pseudoscalars. In the flavour basis, $P_1^0$ couples only with the SM while $P_2^0$ couples only with DM. Their mass mixing induces effective interactions between the SM and the dark matter sector.}

\begin{figure}
\begin{center}
\includegraphics[width=0.43\textwidth]{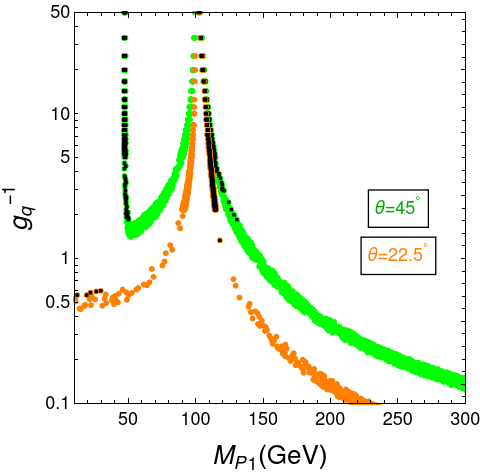}
\includegraphics[width=0.43\textwidth]{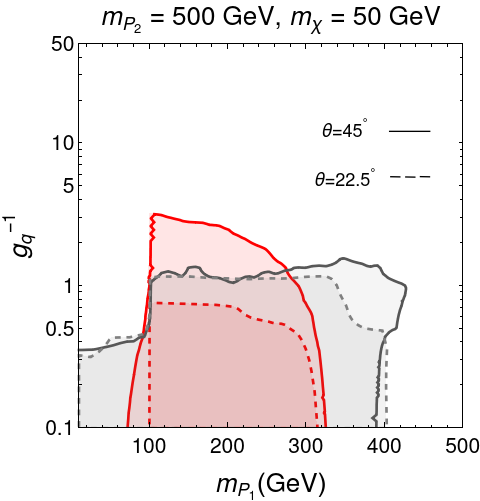}
\caption{\label{fig:kap} \textcolor{Black}{Cosmology (left) and LHC (right) constraints on the model for two representative values of $\theta$, taken to be $\pi/8$ and $\pi/4$. We make use of the same colour code as in figure~\ref {fig:yx}.} } 
  \end{center}
  \end{figure}

\textcolor{Black}{In the no-mixing scenario ($\theta=0$), dark matter does not interact with the SM. Standard freeze-out cannot thus occur. In Sections~\ref{sec:second-mediator} and \ref{sec:dark-matter-mass}, we have considered $\theta = \pi/4$, which corresponds to a maximal-mixing scenario. In order to analyse the effect of varying $\theta$, we consider a scenario where $\theta = \pi/8$, this choice lying mid-way between the no-mixing and maximal-mixing cases. The constraints that we obtain on this scenario are shown in figure~\ref{fig:kap}. The annihilation cross section for $\chi \chi \to$ SM SM scales as $g_q^2 y_\chi^2 \sin^2(2\theta)$, which can be re-written as $g_q^2 {y_\chi^\prime}^2$ with $y_\chi^\prime = y_\chi \sin(2 \theta)$. Therefore, the impact of $\theta$ variations on cosmological constraints is similar to that originating from $y_\chi$ variations. Namely, decreasing $\theta$ from $\pi/4$ to $\pi/8$ reduces the effective coupling $y_\chi^\prime$. Consequently, the range of values of $g_q$ allowed by the relic density constraint increases.}

\textcolor{Black}{The LHC constraints are also sensitive to a change in $\theta$, unlike to a change in the DM Yukawa coupling studied in section~\ref{sec:yx}. With the decrease in $\theta$, the $P_2$ production cross section decreases significantly, thus relaxing any bounds that could stem from mono-Higgs searches. Mono-jet constraints are not much affected when they are dominated by $P_1$ exchanges. However when $P_2$ starts to dominate then the constraints become less stringent, as for when $\theta = \pi/8$. This is illustrated in figure~\ref{fig:kap} at large values of $m_{P_1}$.}

\textcolor{Black}{Consequently, the most stringent constraints on the model are found in the maximal-mixing scenario, as argued in previous sections.}

%%%%%%%%%%%%%%%%%%%%%%%%%%%%%%%%%%%%%%%%%%%%%%%%%%%%%%%%%%%%%%%%%%%%%%%%%%%%%%%%%%%%%%%%%%%%
\subsection{\textcolor{Black}{The impact of the trilinear couplings}}
%%%%%%%%%%%%%%%%%%%%%%%%%%%%%%%%%%%%%%%%%%%%%%%%%%%%%%%%%%%%%%%%%%%%%%%%%%%%%%%%%%%%%%%%%%%%%

\begin{figure}
\begin{center}
\includegraphics[width=0.45\textwidth]{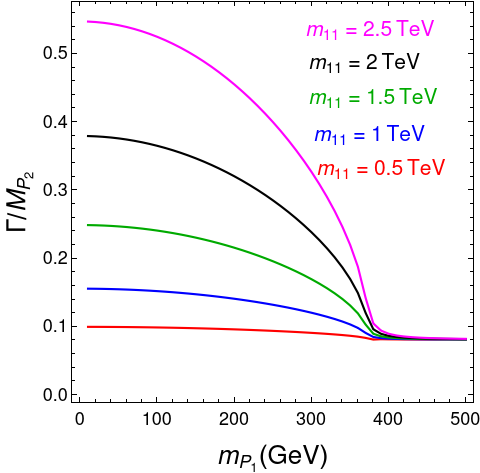}
\includegraphics[width=0.43\textwidth]{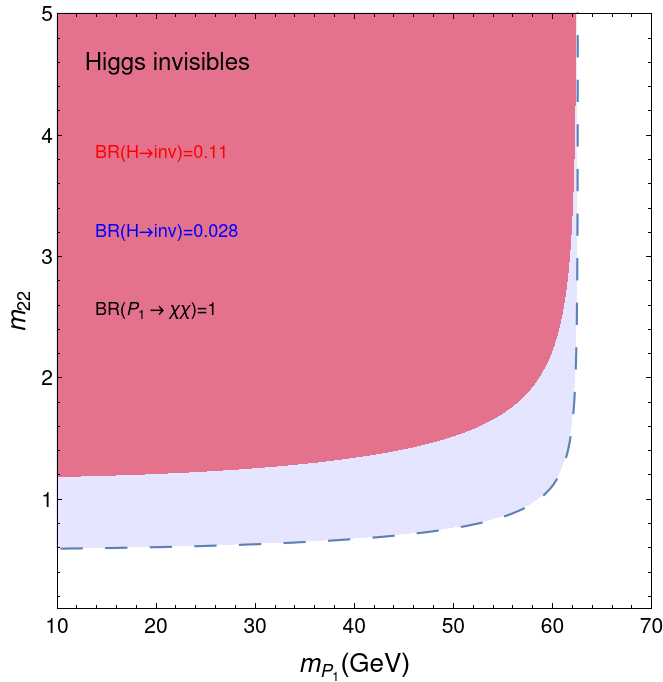}
\caption{\label{fig:width}  {\it Left panel}: dependence of $\Gamma(P_2)/M_{P_2}$ on the lighter pseudoscalar mass for different values of $m_{11}$. {\it  Right panel}: constraints on $m_{22}$ originating from the upper limits on the Higgs-boson invisible width. The shaded regions blue (red) are ruled out by the current (projected at HL-LHC) constraints on the Higgs to invisible branching ratio. } 
  \end{center}
  \end{figure}

\begin{figure}
\begin{center}
\includegraphics[width=0.45\textwidth]{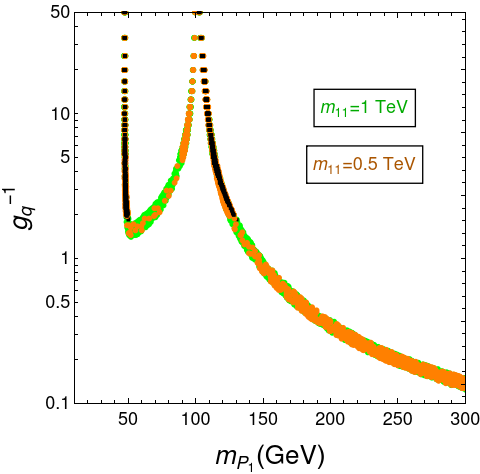}
\includegraphics[width=0.45\textwidth]{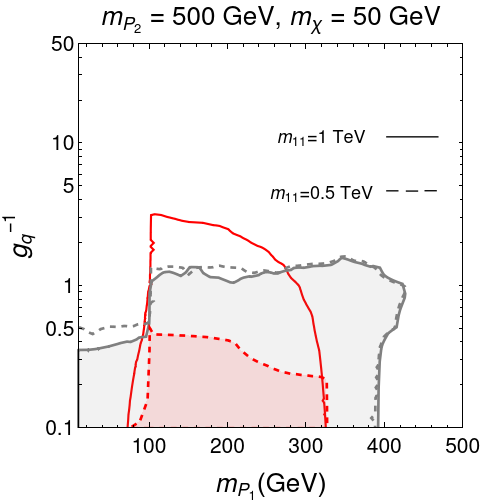}
\caption{\label{fig:m11var} \textcolor{Black}{Cosmology (left) and LHC (right) constraints on the model for two representative values of $m_{11}$, taken to be 1~GeV and 1~TeV. We make use of the same colour code as in figure~\ref {fig:yx}.}} 
 \end{center}
 \end{figure}

In this section we discuss the impact of varying the three trilinear couplings, namely $m_{11}$, $m_{22}$ and $m_{33}$. Amongst these, $m_{11}$ directly affects mono-Higgs production and directly enters the $P_2$ decay width. Its value needs therefore to be carefully chosen. From figure~\ref{fig:width}, it can be seen that $m_{11}$ larger than 1~TeV leads to broad-width scenarios. Since we decided to focus on narrow width scenarios, this therefore justifies our choice of fixing $m_{11}$ to 1~TeV, as performed in the previous sections. Furthermore fixing $m_{11}$ to its maximally allowed value also leads to the most stringent set of constraints that could originate from mono-Higgs searches. Nevertheless, we have studied the impact of reducing $m_{11}$ to  500 GeV, a value which also leads to a branching ratio of 100\% for the $P_2 \to P_1 h_1$ decay, when it is kinematically accessible. We found hardly any impact on both the cosmological and collider constraints. On the contrary, for much lower values like $m_{11}=1$~GeV, mono-Higgs constraints become milder (due to the corresponding cross section being reduced), whereas mono-jet and cosmological constraints are not affected. This is illustrated in figure~\ref{fig:m11var}.

The two other trilinear couplings, $m_{22}$ and $m_{33}$, can be constrained using bounds on the Higgs-boson branching ratio to invisible systems, provided that $M_P{_1} \leq m_H/2$ ($M_P{_2} \leq m_H/2$). The range of $m_{22}$ values allowed by the current constraint BR$_{\rm inv}<11\%$~\cite{ATLAS:2020kdi} corresponds to the region below the dashed line in the right panel of figure~\ref{fig:width}. Here, a 100$\%$ branching ratio of the pseudoscalar to dark matter is assumed. The improvement that can be expected from HL-LHC, leading to BR$_{\rm inv}<2.8\%$~\cite{CMS:2017cwx} is also displayed. The values of these trilinear couplings do not affect any other constraints.

%%%%%%%%%%%%%%%%%%%%%%%%%%%%%%%%%%%%%%%%%%%%%%%%%%%%%%%%%%%%%%%%%%%%%%%%%%%%%%%%%%%%%%%%%%%%%%%
\subsection{Future constraints and challenges at the HL-LHC}
%%%%%%%%%%%%%%%%%%%%%%%%%%%%%%%%%%%%%%%%%%%%%%%%%%%%%%%%%%%%%%%%%%%%%%%%%%%%%%%%%%%%%%%%%%%%%%%

%%%%%%%%%%%%%%%%%%%%%%%%%%%%%%%%%%%%%%%%%%%%%%%%%%%%%%%%%%%%%%%%%%%%%%%%%%%%%%%%%%%%%%%%%%%%%%%
\begin{figure}
\begin{center}
\includegraphics[width=0.43\textwidth]{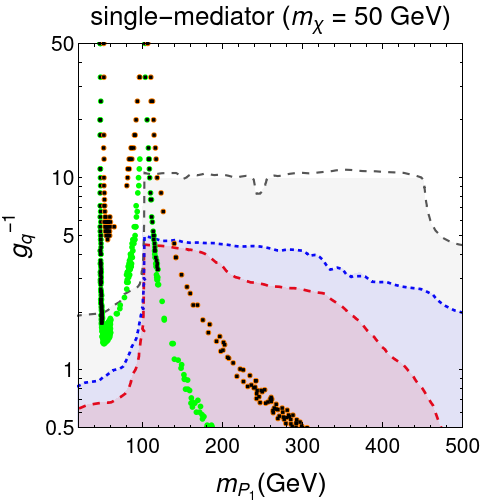}
\includegraphics[width=0.43\textwidth]{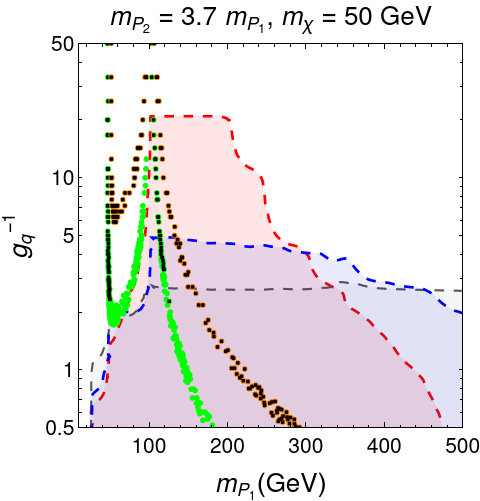} 
\includegraphics[width=0.43\textwidth]{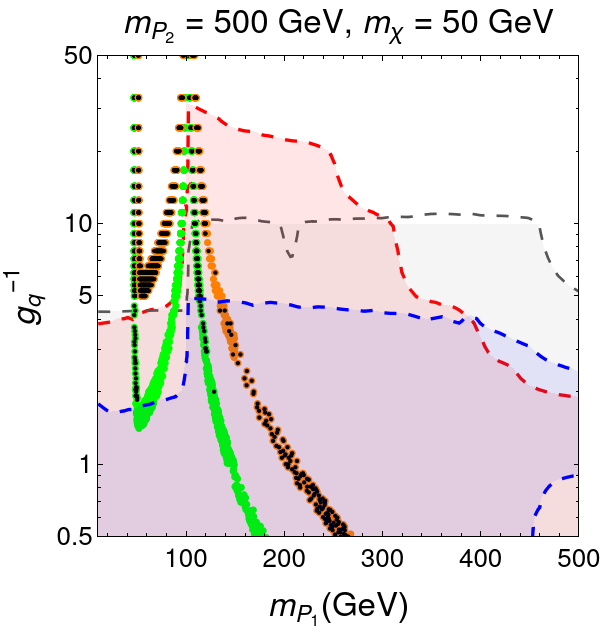}
\includegraphics[width=0.43\textwidth]{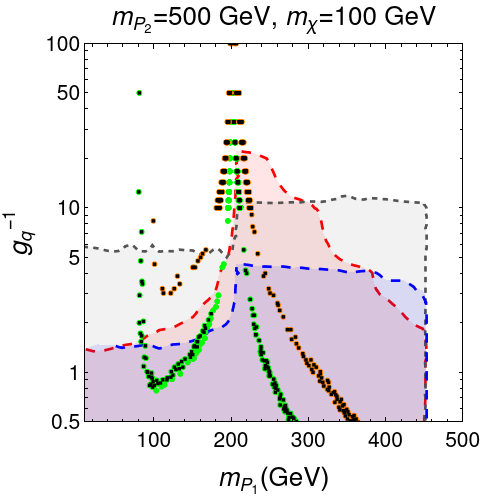}
  \caption{\label{fig:future}Constraints on the model that originate from dark matter relic density and indirect detection, shown together with HL-LHC expectations for a luminosity of 3~ab$^{-1}$. The results are shown in the plane $\left( m_{P_1}, g_q^{-1}\right)$ for the four scenarios discussed in the rest of section~\ref{sec:constraints}. The points appearing in light green are associated with a relic density satisfying $\Omega h^2 = 0.12 \pm 10 \%$ (assuming standard cosmology), whereas the dark green points are additionally allowed by indirect detection results. Points shown in orange correspond to cosmological scenarios in which there was a period of matter domination in the early universe with a dilution factor of 10. The shaded regions represent 95\% confidence level expected exclusions from mono-jet (grey), mono-Higgs (red) and $t\bar{t}+\slashed{E}_T$ (blue) analyses in 3~ab$^{-1}$ of HL-LHC data, as extrapolated from the searches of refs.~\cite{ATLAS:2017bfj, ATLAS:2017uis, CMS:2017jrd}.} 
  \end{center}
  \end{figure}
%%%%%%%%%%%%%%%%%%%%%%%%%%%%%%%%%%%%%%%%%%%%%%%%%%%%%%%%%%%%%%%%%%%%%%%%%%%%%%%%%%%%%%%%%%%%%%%

In order to assess how well multi-pseudoscalar mediated dark matter models can be probed in future LHC operation runs, we extrapolate current constraints to an integrated luminosity of 3~ab$^{-1}$. In our naive extrapolation procedure, we rescale the background expectation and signal predictions to match the HL-LHC luminosity, and assume identical relative uncertainties on the background. The HL-LHC sensitivity derived in this way is displayed in figure~\ref{fig:future} for the four classes of benchmark scenarios considered. After accounting for all cosmology and collider bounds, only funnel regions end up to be allowed by (future expected) data. Such a conclusion is similar to the one obtained for Higgs-portal mediated dark matter scenarios, with a subtle difference that in a multi-pseudoscalar setup the HL-LHC is competitive enough to constrain parts of the allowed regions of the parameter space. Scenarios featuring coupling modifiers of ${\cal O}(0.1)$ or smaller are indeed the only ones expected to survive without being tested after the future high-luminosity phase of the LHC.

All the results presented so far assume standard cosmology. We now examine the fate of the model considered in the light of non-standard early universe cosmology. Experimentally we know that the universe was radiation dominated at the Big Bang Nucleosynthesis epoch. Prior to that and in the absence of any relevant information, a radiation-dominated universe is usually considered as well, as such an assumption leads to a more predictive setup. However, departures from this condition are not excluded, as for example when there is a period of early matter domination (see, for example, ref.~\cite{Gelmini:2006pw}). Such a modification in the cosmology would cause dilution in the number densities, yielding cosmological constraints that are potentially relaxed quite a bit. In figure~\ref{fig:future}, we present how a simple dilution factor of 10 can modify the cosmological bounds on the model. Favoured scenarios exist and are quite challenging to be probed at the HL-LHC. In this way, whereas the LHC and the HL-LHC may be sensitive to parameter space regions favoured by standard cosmology, the situation changes when non-standard effects are in order. In particular, allowed parameter space regions close to the resonance regime ($m_{P_1}\sim 2 m_\chi$) remain largely untestable at the HL-LHC by virtue of too small new physics couplings of the dark sector to the SM.

%%%%%%%%%%%%%%%%%%%%%%%%%%%%%%%%%%%%%%%%%%%%%%%%%%%%%%%%%%%%%%%%%%%%%%%%%%%%%%%%%%%%%%%%%%%%%%%%%%%%%%%%%%%%%%%
\section{Conclusions}
\label{sec:conclusions}
%%%%%%%%%%%%%%%%%%%%%%%%%%%%%%%%%%%%%%%%%%%%%%%%%%%%%%%%%%%%%%%%%%%%%%%%%%%%%%%%%%%%%%%%%%%%%%%%%%%%%%%%%%%%%%%

In this work, we focus on less-simplified models for dark matter and their associated phenomenology. We classify such setups based on the number of additional mediators and dark matter particles relevant for LHC energy scales. The classification is simple and allows us to retain the main features of the usual simplified models for dark matter, that are quite popular in serving as frameworks to jointly examine cosmological and collider searches. However, they have been taken over by less-simplified models owing to the greater phenomenological reach and theoretical consistency of the latter, even if in less-simplified models we get plagued with larger sets of free parameters so that any analysis becomes less transparent. We are therefore largely dependent on the details of the UV physics to reduce the complexity of the model. This context thus necessitates the need for finding pathways for model-independent searches.

We establish this by using a simple example of two-pseudoscalar mediated simplified models for dark matter. The pseudoscalar mediators are popular candidates to relate the SM and DM sectors, as they ease to avoid direct detection constraints on the model parameters. Having two pseudoscalar mediators consists, on the other hand, of a natural configuration when one aims to introduce new physics interactions in a gauge-invariant way on top of a standard simplified DM model. To conduct our analysis of the two-mediator model, we make certain simplifying assumptions. We require the lightest Higgs-boson interactions to be SM-like, that the interactions of the pseudoscalars satisfy a minimally flavour-violating structure and are universal across all quark species. Moreover, the strength of the Yukawa coupling of the mediators with DM are held fixed, as this parameter does not contribute much to the collider phenomenology, and the scalar trilinear couplings are estimated from Higgs physics considerations. With these assumptions, the presentation of the multi-mediator analysis becomes simple, and depends only on four parameters.

We consider experimental constraints originating from the DM relic density, indirect detection, and mono-$X$ searches at the LHC. In particular, we focus on two classes of scenarios to tackle two separate issues. First, we aim to factorise the effect of having a not too heavy second pseudoscalar in the spectrum, or in other words to be able to distinguish between a single-mediator and a two-mediator setup even when the second mediator does not contribute to standard mono-jet signals. We find that mono-Higgs probes can consist of new handles on regions of the model parameter space as soon as the second mediator is present in the low-energy part of the spectrum. Second, we assess the impact of the dark matter mass on the constraints for a given second mediator mass. Here, for larger values of the DM mass constraints from indirect detection relax. Furthermore, we show that in all cases, the LHC can quite stringently restrict the allowed regions of the parameter space with a luminosity of 3 ab$^{-1}$, so that only tiny funnel-like regions near $m_{P_1} \approx m_\chi/2$ survive. Those constraints, however, get weaker when non-standard cosmology, before dark matter freeze-out, is assumed. Taking the example of matter domination in the early universe, a simple dilution factor of 10 is found to be sufficient to leave large parts of the parameter space unconstrained.
%%%%%%%%%%%%%%%%%%%%%%%%%%%%%%%%%%%%%%%%%%%%%%%%%%%%%%%%%%%%%%%%%%%%%%%%%%%%%%%%%%%%%%%%%%%%%%%%%%%%%%%%%%%%%%%%%%%%
%%%%%%%%%%%%%%%%%%%%%%%%%%%%%%%%%%%%%%%%%%%%%%%%%%%%%%%%%%%%%%%%%%%%%%%%%%%%%%%
\section*{Acknowledgments}
%%%%%%%%%%%%%%%%%%%%%%%%%%%%%%%%%%%%%%%%%%%%%%%%%%%%%%%%%%%%%%%%%%%%%%%%%%%%%%% 
The authors would like to thank Jack Y. Araz and Emanuele Re for useful discussions and important checks. We would further like to acknowledge IACS, Kolkata and Abhaya Kumar Swain for computational support. D.B. would like to acknowledge the support received from the Indo-French LIA grant which she used to travel to LAPTh, Annecy, where this project commenced, and Kasinath Das, Siddharth Dwivedi, and Tuhin Roy for useful correspondences. 
%%%%%%%%%%%%%%%%%%%%%%%%%%%%%%%%%%%%%%%%%%%%%%%%%%%%%%%%%%%%%%%%%%%%%%%%%%%%%%%%% 
%%%%%%%%%%%%%%%%%%%%%%%%%%%%%%%%%%%%%%%%%%%%%%%%%%%%%%%%%%%%%%%%%%%%%%%%%%%%%%%%%%%%%%%%%%%%%%%%%%%%%%%%%%%%%%%%%%
\providecommand{\href}[2]{#2}\begingroup\raggedright\endgroup

%%%%%%%%%%%%%%%%%%%%%%%%%%%%%%%%%%%%%%%%%%%%%%%%%%%%%%%%%%%%%%%%%
\end{document}